\documentstyle[12pt]{article}
\def\journal{\topmargin .3in	\oddsidemargin .5in
	\headheight 0pt	\headsep 0pt
	\textwidth 5.625in 
	\textheight 8.25in 
	\marginparwidth 1.5in
	\parindent 2em
	\parskip .5ex plus .1ex		\jot = 1.5ex}

 \catcode`\@=11

\@addtoreset{equation}{section}

\def\theequation{\thesection.\arabic{equation}}

\newtoks\@stequation

\def\subequations{\refstepcounter{equation}%
  \edef\@savedequation{\the\c@equation}%
  \@stequation=\expandafter{\theequation}
  \edef\@savedtheequation{\the\@stequation}
  \edef\oldtheequation{\theequation}%
  \setcounter{equation}{0}%
  \def\theequation{\oldtheequation\alph{equation}}}

\def\endsubequations{\setcounter{equation}{\@savedequation}%
  \@stequation=\expandafter{\@savedtheequation}%
  \edef\theequation{\the\@stequation}\global\@ignoretrue}

\journal

\catcode`\@=11
\def\marginnote#1{}
\catcode`\@=11
\def\section{\@startsection {section}{1}{0pt}{-3.5ex plus -1ex minus
 -.2ex}{2.3ex plus .2ex}{\raggedright\large\bf}}
\catcode`\@=12
\newskip\humongous \humongous=0pt plus 1000pt minus 1000pt
\def\caja{\mathsurround=0pt}
\def\eqalign#1{\,\vcenter{\openup1\jot \caja
	\ialign{\strut \hfil$\displaystyle{##}$&$
	\displaystyle{{}##}$\hfil\crcr#1\crcr}}\,}
\newif\ifdtup

\def\Q{{\mathchoice
{\setbox0=\hbox{$\displaystyle\rm Q$}\hbox{\raise 0.15\ht0\hbox to0pt
{\kern0.4\wd0\vrule height0.8\ht0\hss}\box0}}
{\setbox0=\hbox{$\textstyle\rm Q$}\hbox{\raise 0.15\ht0\hbox to0pt
{\kern0.4\wd0\vrule height0.8\ht0\hss}\box0}}
{\setbox0=\hbox{$\scriptstyle\rm Q$}\hbox{\raise 0.15\ht0\hbox to0pt
{\kern0.4\wd0\vrule height0.7\ht0\hss}\box0}}
{\setbox0=\hbox{$\scriptscriptstyle\rm Q$}\hbox{\raise 0.15\ht0\hbox to0pt
{\kern0.4\wd0\vrule height0.7\ht0\hss}\box0}}}}
\def\C{{\mathchoice
{\setbox0=\hbox{$\displaystyle\rm C$}\hbox{\hbox to0pt
{\kern0.4\wd0\vrule height0.9\ht0\hss}\box0}}
{\setbox0=\hbox{$\textstyle\rm C$}\hbox{\hbox to0pt
{\kern0.4\wd0\vrule height0.9\ht0\hss}\box0}}
{\setbox0=\hbox{$\scriptstyle\rm C$}\hbox{\hbox to0pt
{\kern0.4\wd0\vrule height0.9\ht0\hss}\box0}}
{\setbox0=\hbox{$\scriptscriptstyle\rm C$}\hbox{\hbox to0pt
{\kern0.4\wd0\vrule height0.9\ht0\hss}\box0}}}}

\font\fivesans=cmss10 at 4.61pt
\font\sevensans=cmss10 at 6.81pt
\font\tensans=cmss10
\newfam\sansfam
\textfont\sansfam=\tensans\scriptfont\sansfam=\sevensans\scriptscriptfont
\sansfam=\fivesans
\def\sans{\fam\sansfam\tensans}
\def\Z{{\mathchoice
{\hbox{$\sans\textstyle Z\kern-0.4em Z$}}
{\hbox{$\sans\textstyle Z\kern-0.4em Z$}}
{\hbox{$\sans\scriptstyle Z\kern-0.3em Z$}}
{\hbox{$\sans\scriptscriptstyle Z\kern-0.2em Z$}}}}

\mathchardef\endbar="375

\def\ceilfill{$\raise3pt\hbox{$\mathsurround=0pt\mathord\endbar$}
  \mkern-2mu \xleaders\hbox{$\mkern-5mu
  \mathord-\mkern-5mu$}\hfill\mkern-7mu
  \raise3pt\hbox{$\mathsurround=0pt\mathord\endbar$}$}

\def\floorfill{$\raise9pt\hbox{$\mathsurround=0pt\mathord\endbar$}
  \mkern-2mu \xleaders\hbox{$\mkern-5mu
  \mathord-\mkern-5mu$}\hfill\mkern-7mu
  \raise9pt\hbox{$\mathsurround=0pt\mathord\endbar$}$}

\def\overcontract#1{\mathop{\vbox{\ialign{##\crcr\noalign{\kern3pt}
  \ceilfill\hskip6pt\crcr\noalign{\kern3pt\nointerlineskip}
  $\hfil\displaystyle{#1}\hfil$\crcr}}}}

\def\undercontract#1{\mathop{\vtop{\ialign{##\crcr
  $\hfil\displaystyle{#1}\hfil$\crcr\noalign{\kern3pt\nointerlineskip}
  \floorfill\hskip6pt\crcr\noalign{\kern3pt}}}}}
\def\a{\alpha}
\def\b{\beta}

\def\d{\delta}
\def\e{\epsilon}
\def\m{\mu}
\def\n{\nu}
\def\t{\tau}
\def\p{\pi}

\def\r{\rho}
\def\th{\theta}
\def\s{\sigma}
\def\l{\lambda}
\def\o{\omega}
\def\k{\kappa}

\def\x{\xi}

\def\L{\Lambda}

\def\Fi{\Phi}
\def\O{\Omega}
\def\D{\Delta}
\def\z{\zeta}
\def\bz{\bar{\zeta}}

\def\tR{\hat{R}}
\def\bs{\bar{s}}

\def\fb{\bar{f}}

\def\bA{\bar{A}}
\def\bB{\bar{B}}

\def\ba{\bar{\alpha}}
\def\tJ{\tilde J}

\def\tc{\tilde{c}}
\def\ts{\tilde{s}}

\def\C{{\cal C}}

\def\pa{\partial}
\def\bpa{\bar{\partial}}

\def\ra{\rightarrow}

\def\ti{\times}

\def\xx{\hbox{ }^*_*}

\def\rd{{\rm d}}

\def\ed{\end{document}}

\def\be{\begin{equation}}
\def\ee{\end{equation}}
\def\bs{\begin{subequations}}
\def\es{\end{subequations}}
\def\ben{\begin{enumerate}}
\def\een{\end{enumerate}}
\def\ia{\item[a)]}
\def\ib{\item[b)]}

\begin{document}

\begin{titlepage}
\begin{center}
April 1994     \hfill        \hfill CPTH-A299.0494 \\
 hep-th/9403191 \hfill $\;$     \\

\vskip .4in

{\large \bf Plane Gravitational Waves in String Theory}
\footnote{Research supported in part by the EEC contracts SC1--CT92--0792
and CHRX-CT93-0340.
The work of NO was supported in part by a
S\'ejour Scientifique de Longue Dur\'ee of the Minist\`ere des Affaires
Etrang\`eres.}

\vskip .3in
I. Antoniadis
\footnote{e-mail: ANTONIAD@ORPHEE.POLYTECHNIQUE.FR}
 and
N.A. Obers
\footnote{e-mail: OBERS@ORPHEE.POLYTECHNIQUE.FR }
\vskip .1in
{\em Centre de Physique Th\'eorique \footnote{Laboratoire Propre du
CNRS UPR A.0014} \\
Ecole Polytechnique  \\
F-91128 Palaiseau \\
France}
\vskip .3in

\end{center}
\vskip .1in
\begin{abstract}
We analyze the coset model $( E_2^c \ti E_2^c)/E_2^c$ and construct a
class of exact string vacua which
describe plane gravitational
waves and their duals, generalizing
the plane wave background found by Nappi
and Witten.
In particular, the vector gauging describes a two-parameter family of singular
geometries
with two isometries, which is dual to plane gravitational waves.
In addition, there is a mixed vector-axial gauging which describes a
one-parameter family of plane waves with five isometries.
These two backgrounds are related by a duality transformation which generalizes
the known axial-vector duality for abelian subgroups.

\end{abstract}
\end{titlepage}
\renewcommand{\thepage}{\arabic{page}}
\setcounter{page}{1}
\setcounter{footnote}{1}
\section{Introduction}
The search of exact non-trivial background solutions in string theory plays
an important role in understanding the structure of space time at short
distances and how string dynamics modifies gravitational interactions.
In particular, one is interested to discover how string theory affects space
time singularities that are present in many solutions of classical general
relativity, such as black holes. On the one hand one would like to find
string generalizations of known non-trivial solutions of Einstein's
equations, while on the other hand one would like to determine the
corresponding conformal field theories which allows
in principle to study the string excitations and their interactions
around these backgrounds.

Although this background identification is useful to establish the connection
of
string solutions with known spaces in general relativity, its validity is
restricted only to the region of small curvatures where string interactions can
be neglected. An important tool for relating different background solutions
which are valid in separate kinematic regions of the same string vacuum is
provided by duality transformations (see e.g. \cite{gpr} for a review).
In this way, one may relate spaces which
look very different from a geometric and topological point of view, in
particular exhibiting different singularity structure, though originating from
the same conformal field theory (CFT). Duality was shown to be an exact
symmetry of string theory in the case of compact spaces \cite{dua} as well as
in
non-compact abelian coset models \cite{gk}, a property which is believed to be
valid generally.

An interesting class of solutions of Einstein's equations consists of plane
gravitational waves \cite{bm,es}.
It was recently shown \cite{nw} that the most symmetric gravitational wave
(with seven isometries)
can be extended to an exact string background, described by a
Wess-Zumino-Witten
(WZW) model on the non-semi-simple group $E_2^c$ which is the central extension
of the Euclidean group in two dimensions. Furthermore, it was shown that any
plane gravitational wave can be extended to an exact string background
\cite{kt},
although the underlying conformal field theories remain to be discovered.
Finally, several generalizations \cite{gen} of the $E_2^c$ WZW model
and various gaugings \cite{sf1,kk}
on non-semi-simple groups were considered, while in Ref.\cite{kk} the
representations  of affine $E_2^c$ were constructed in terms of free fields.

In this work, we identify a CFT which describes a large class of plane
gravitational waves and their duals, depending on continuous parameters. It is
based on the gauged WZW model
$( E_2^c \ti E_2^c)/E_2^c$ which contains an arbitrary parameter corresponding
to the continuous embedding of the subgroup $H=E_2^c$ in the product group
$E_2^c \ti E_2^c$. Moreover, we find two inequivalent ways of gauging:
\ben
\ia
The vector gauging depends on an additional continuous parameter and it
describes a singular geometry with two isometries, which is dual to plane
gravitational waves.
\ib
A mixed vector-axial type gauging, based on the existence of a non-trivial
outer automorphism of the subgroup $H=E_2^c$, which describes a one-parameter
family of plane gravitational waves with 5 isometries.
\een
These two geometries are shown to be dual to the same background which
corresponds again to a class of plane gravitational waves. The latter, however,
depends on two parameters and has 6 Killing symmetries. This duality
generalizes the known axial-vector duality for abelian subgroups.

The paper is organized as follows. In Section 2, we present a general
discussion of string backgrounds describing plane gravitational waves which can
be classified according to their number of Killing symmetries. We also compute
their duals with respect to the abelian non-null isometries. In Section 3, we
review the WZW model on
$E_2^c$, which describes the most symmetric case of plane waves with 7
isometries. In Section 4, we show how outer automorphisms of the subgroup $H$
generate inequivalent $G/H$ coset constructions, and we derive the
corresponding general formulae for the $( E_2^c \ti E_2^c)/E_2^c$
coset model. The
resulting backgrounds emerging from the vector- and vector-axial-gauged models
are worked out and studied in Sections 5 and 6, respectively. In Section 7, the
connection between these two backgrounds is established by computing their dual
geometry. Finally, Section 8 contains conclusions and comments for open
directions.

\section{Plane Gravitational Waves}

We start with a discussion of the most general class of conformally invariant
$\s$-models
\be
I = \int \rd^2 z [(G_{\m \n }(x) + B_{\m \n }(x)) \bpa x^{\m} \pa x^{\n}
+ \a ' R^{(2)} \Fi (x) ]
\label{sm} \ee
that correspond to strings propagating in a background of plane gravitational
waves.
Here $G_{\m \n}$ is the background metric, $B_{\m\n}$ the antisymmetric
tensor and $\Fi$ the dilaton.

It was shown in Ref.\cite{kt} (though in a different basis)
that the metric, antisymmetric tensor and dilaton given by
\bs
\be
\rd S^2 = 2 \rd \z \rd \bz  - 2( f(u) \z^2 + \fb (u) \bz^2
+ F(u) \z \bz ) \rd u^2 - 2 \rd u \rd v
\label{gf} \ee
\be
B_{\z \bz} = i b(u)\;\;\;\;, \;\;\;\;  \Fi = - \ln g(u)
\label{bd} \ee
\be
F(u) = {g(u)''  \over g(u)} - \left( { g(u) ' \over g(u) }\right)^2
 + \frac{1}{4}
(b(u)')^2
\label{beq} \ee
\label{pgw} \es
satisfy the one-loop beta functions of conformal invariance, where
$f(u)$ is an arbitrary complex funtion, and $g(u)$ and $b(u)$ are arbitrary
real
functions. Primes denote derivatives with respect to $u$.

The metric (\ref{gf}) describes a plane gravitational wave, and the
antisymme- \linebreak tric tensor and dilaton are chosen such that they
 respect the Killing
symmetries of the gravitational wave. In fact, eq.(\ref{beq}) on $F(u)$
is a consequence of the one-loop beta functions \cite{beta}
\bs
\be
\b^G_{\m\n} =
R_{\m \n} - \frac{1}{4} H_{\m}{}^{\l \s} H_{\n \l \s} + 2 \nabla_{\m}
\nabla_{\n} \Fi=0
\ee
\be
\b^B_{\m \n} =
\nabla_{\l} H^{\l}{}_{\m \n} -2 (\nabla_{\l} \Fi) H^{\l}{}_{\m \n}=0
\ee
\be
\b^\Fi=
{c- 4 \over 3 \a '} + 4 (\nabla \Fi)^2 - 4 \nabla^2 \Fi - R + \frac{1}{12} H^2
=0 \label{betadil} \ee
\label{beta} \es
where $H_{\m \n \l} = 3 \nabla_{[\m} B_{\n \l]}$ is the antisymmetric
tensor field strength and $R_{\m \n}$ is the Ricci tensor. One finds that for
general $F(u)$ all equations are satisfied, except $\b^G_{uu}$, whose vanishing
then results
in eq.(\ref{beq}). Moreover, because all scalar invariants vanish the
central charge deficit $c-4$ in (\ref{betadil}) is zero.

The non-zero components of the Rieman tensor, Ricci tensor,
antisymmetric field strength,
and squared antisymmetric field strength,
for the
the background in (\ref{pgw}) are
\bs
\be
R_{\z u \z u} = 2 f(u) \;\;\;\;,\;\;\;\; R_{\bz u\bz u} = 2\fb(u) \;\;\;\;,
\;\;\;\; R_{\z u \bz u} = F(u)
\;\;\;\;, \;\;\;\; R_{uu} =
2 F(u)
\ee
\be
H_{\z \bz u} = i b(u)' \;\;\;\;, \;\;\;\; H^2_{uu}= 2(b(u)')^{2}
\ee
\label{geoq} \es
and the scalar curvature is obviously zero. Moreover, it was argued
\cite{nw,kt}
that the 1-loop solution is an exact background to all orders in $\a'$,
as one can check that all higher-order contractions of the relevant tensors
vanish.

For general $f(u)$, $g(u)$ and $b(u)$, the geometry in (\ref{pgw}) has
five Killing
vectors, one of which is the null Killing vector
\be
 k^{\m} = (0,0,0,1)
\label{nkv} \ee
which is characteristic for gravitational waves. In eq.(\ref{nkv}), $\m$
refers to $\z,\bz,u,v$, in that order. In fact, the metric (\ref{gf})
is of pure radiation type \cite{es}, which describes a solution of Einstein's
equations with energy momentum tensor
\be T_{\m \n } =  T(u) k_{\m } k_{\n}\;\;\;\;,\;\;\;\;  k^{\m} k_{\m} = 0
\;\;\;\;. \ee
Moreover, the null Killing vector is covariantly constant, which defines the
subset of plane-fronted gravitational waves, discovered by Brinkmann
\cite{bm}, which
in turn includes the plane wave solution (\ref{gf}) as a special case.
Its additional four space-like Killing vectors are given by
\be
\xi^{\m} =(\l,\bar{\l} ,0, \z  \bar{\l}' + \bz \l' )
\label{kv} \ee
where $\l(u)$ satisfies the differential equation
$ \l '' + 2 \fb (u) \bar{\l} ' + F(u) \l =0 $, which gives rise to four
integration constants in (\ref{kv}).

Among the five Killing vectors one can find
a subset of three commuting Killing vectors,
one null and
two space-like, wich are manifest in the following alternate form of the
background
\bs
\be
\rd S^2  = g_{mn}(u)  \rd x^n \rd x^n - 2 \rd u \rd v '
\;\;\;\;, \;\;\;\; g_{mn} = 2 \ba_{(m } \a_{n) }
\ee
\be
B_{1 2 } = i \int \rd u [(\a_1 \ba_2 - \ba_1 \a_2)  b(u)']
\;\;\;\;  , \;\;\;\; \Fi(u) = -\ln (g(u))
\label{ast} \ee
\label{f2} \es
where we have made the coordinate transformation
\bs
\be
\z = \a_m(u) x^m \;\;\;\;, \;\;\;\; v = v' + \frac{1}{4} g_{mn} (u)' x^m x^n
\ee
\be
{\rm Re} [ \ba_{(m} \a_{n)}'' + 2 f(u) \a_m \a_n + F(u) \ba_{m} \a_{n} ] = 0
\label{aeq} \ee
\label{bt} \es
with $\a_m(u),\;m=1,2$ complex functions satisfying (\ref{aeq}). This
coordinate transformation also generates non zero components $B_{1u}$ and
$B_{2u}$, which were gauged away by a gauge transformation
$B_{\m \n} \ra B_{\m \n} + \pa_{\m} \L_{\n} - \pa_{\n} \L_{\m}$.

A special case of the metric is obtained when $f=0$. In this case, we have
at least six Killing vectors, the null Killing vector (\ref{nkv}) and five
space-like Killing vectors wich are given by
\bs
\be
\x^{\m} = ( h,\pm h  , 0,(\bz \pm \z)  h' )
\;\;\;\;, \;\;\;\; h'' + F h =0
\label{skv} \ee
\be
\x^{\m}= (\z ,-\bz, 0,0)
\label{rkv} \ee
\label{kv2} \es
where $h(u)$ contains two integration constants.
Note that compared to the general case, we have the extra ``rotational''
symmetry (\ref{rkv}).
If in addition $F$ is independent of $u$, so that
$m^{\m} = (0,0,1,0)$ is also a Killing vector, we arrive at the
most symmetric (non-trivial) plane gravitational wave, with seven isometries.
The background found by Nappi and Witten \cite{nw}
for the $E_2^c $ WZW model corresponds
to this case (see Section 3).

For use below, we give here the basis in which the rotational symmetry is
manifest. First, we perform the basis transformation in (\ref{bt}), with
$\a_1 = -i \a_2= h(u)/\sqrt{2}$, where $h(u)$ is a solution of the differential
equation (\ref{skv}). Subsequently, we define
polar coordinates $x^1 = r \cos \th, \;x^2 =r \sin \th$, and
find
\bs
\be
\rd S^2 =  h(u)^2 (\rd r^2 + r^2 \rd \th^2) - 2 \rd u \rd v'
\ee
\be
B_{r \th} =  r \int \rd u [h(u)^2  b(u)'] \;\;\;\; , \;\;\;\; \Fi = -\ln(g(u))
\ee
\label{f3} \es
showing the abelian isometry in the $\th$-direction.

The various gravitational waves backgrounds,
classified according to the number of Killing symmetries they possess,
are summarized in Table 1. Here $f,\;F$ without arguments stand for
$u$-independent
constants.
\newpage
\begin{center}
Table 1. Number of Killing symmetries for plane gravitational waves
\newline
\vskip .3cm
\begin{tabular}{|r || c | c | c ||} \hline
& $ f(u) $ & $ F(u) $ & dim(Killing sym.)  \\ \hline \hline
I & $ 0 $ & $ F \neq 0 $ & 7   \\ \hline
II & $ 0 $ & $ F(u)  $ & 6   \\ \hline
III & $ f\neq 0  $ & $ F $ & 6   \\ \hline
IV & $ f(u) $ & $ F(u)  $ & 5   \\ \hline
\end{tabular}
\end{center}

We  next turn to the study of the  background geometries that can
be obtained from duality transformations of the plane gravitational wave
background. We will restrict to the following two cases:
\ben
\ia $O(2,2)$ duality corresponding to the two space-like Killing symmetries
that are manifest in (\ref{f2}).
\ib $O(1,1)$ duality in the case of $f=0$, corresponding to the (rotational)
space-like Killing vector manifest in (\ref{f3}).
\een
These possibilities were also discussed in Ref.\cite{kt}, and, in particular,
applied
to the plane gravitational wave of Ref.\cite{nw}.
Moreover, a more general possibility
of including the null-isometry (combined with other non-null isometries
as to avoid singularities
in the duality inversion) was considered to show that the Nappi and
Witten plane wave is dual to flat space \cite{kk,kt}
with constant antisymmetric tensor
and dilaton, by an $O(3,3)$ rotation. Finally, we note that in the cases
I and III the
$u$-isometry can be incorporated in the duality transformations, though we
will not work this out explicitly.

For the case of $d$ abelian isometries in a $D$-dimensional background
geometry, the duality transformations read \cite{gpr}
\bs
\be
Q'_{ij}= Q_{ij} - Q_{i a} (Q^{-1})^{ab} Q_{b j}
\;\;\;\;, \;\;\;\;
Q'_{ab} = (Q^{-1})_{ab}
\ee
\be
Q'_{ia} = - Q_{ib} (Q^{-1})^b{}_{a}
\;\;\;\;,\;\;\;\;
Q'_{ai} =  (Q^{-1})_{a}{}^b Q_{b i}
\ee
\be
\Fi' = \Fi - \frac{1}{2} \ln ({\rm det}(G_{ab}))
\ee
\be
Q_{\m \n } = G_{\m \n } + B_{\m \n } =\left( \matrix{
Q_{ij}(x^i) & Q_{ia} (x^i) \cr
Q_{ai}(x^i) & Q_{ab} (x^i) \cr } \right)
\ee
\label{dual} \es
where the matrix $Q$ and the dilaton $\Fi$ are independent of the
$d$ coordinates $x^a, a=1,\ldots d$, and the remaining $D-d$ coordinates are
labelled by $x^i$.

First, it is not difficult to see using eq.(\ref{dual}) that the
dual of
(\ref{f2}) with respect to either the two commuting space-like isometries, or
with respect to any linear combination of these, is again a gravitational
wave.

On the other hand, the dual geometry with respect to the $\th$-isometry in
(\ref{f3})
is
\bs
\be
\rd S^2 =  h(u)^2 (1 + l(u)^2) \rd r^2 -  \frac{ 2 l(u) }{r} \rd r \rd \th
+ \frac{1}{ r^2 h(u)^2} \rd \th^2 - 2 \rd u \rd v
\ee
\be
l(u) =\frac{1}{h(u)^2} \int \rd u [h(u)^2 b(u)']
\label{bkrel} \ee
\be
B_{\m \n } =0 \;\;\;\;, \;\;\;\; \Fi = -\ln(g(u)) - \ln ( r h(u))
\ee
\label{geod} \es
which is clearly not anymore of the type (\ref{pgw}), and corresponds to
a curved background with
Ricci tensor and scalar curvature,
\bs
\be
R_{rr} = -2 { 1 +l^2 \over r^2}
\;\;\;\;, \;\;\;\; R_{r \th} =  { 2 g \over r^3 h^2}
\;\;\;\;, \;\;\;\;
R_{\th \th } = - {2 \over r^4 h^4}
\ee
\be
R_{r u} = - {2(1+l^2)(h'/h) + l l' \over r}
\;\;\;\;, \;\;\;\;
R_{\th u} =  { l' + 2 l (h'/h) \over r^2 h^2}
\ee
\be
R_{ uu } = -2( 1+ l^2) (h'/h)^2 - 2 l l' (h'/h) - \frac{1}{2} (l')^2
\ee
\be
R = - { 4 \over r^2 h^2}
\ee
\label{geoq2} \es
showing singularities at $r=0$. This solution can be viewed as a new
singular solution to Einstein's equations, with non-trivial matter.
We remind the reader that the function $h(u)$ is dependent on $g(u)$ and
$b(u)$,
through eq.(\ref{beq}) and the differential equation in (\ref{skv}).
We have checked that the metric has in general no other Killing symmetries,
besides the manifest space-like and null Killing isometry.

In the remainder of the paper, we will find explicit conformal field
theoretic realizations of the plane wave geometries and their duals, discussed
above.

\section{The WZW Model on ${\bf E_2^c} $}
In this section we review the WZW action
on the non-semi-simple group
$E_2^c$ \cite{nw}, whose algebra is given by
\bs
\be
[J,P_i] = \e_{ij} P_j \;\;\;\;,\;\;\;\;
[P_i,P_j] = \e_{ij} T  \;\;\;\;,\;\;\;\;
[T,J] = [T,P_i] = 0
\ee
\be
[T_a,T_b] = f_{ab}{}^c T_c \;\;\;\;,\;\;\;\;
 \{ T_a\, |\; a =1, \ldots, 4 \} = \{ P_1 , P_2, J, T \}
\ee
\es
which is a central extension of the two-dimensional Poincare algebra.
Although the Killing metric $\eta_{ab} = f_{ac}{}^d f_{bd}{}^c$
is degenerate,
there exists a non-degenerate invariant bilinear form,
\be
\O_{ab} = k \left( \matrix{ 1 & 0 & 0 & 0 \cr
0 & 1 & 0 & 0 \cr 0 & 0 & b & 1 \cr 0 & 0 & 1 & 0 \cr} \right)
 \label{bil} \ee
which is symmetric and satisfies $\O_{ad} f_{bc}{}^d + \O_{bd} f_{ac}{}^d=0$.
Due to these properties, the
 bilinear form can be used to construct the WZW action
\be
I_{WZW}(g)= -\frac{1}{4\p}\int_{\Sigma}  \rd^2 z {\rm Tr}
(g^{-1} \pa g g^{-1} \bpa g)
+ \frac{i}{12\p } \int_{B} \rd^3 z {\rm Tr}( g^{-1} d g)^3
\label{wzw} \ee
by replacing ${\rm Tr}(T_a T_b) $ with $\O_{ab}$.

To evaluate the action explicitly one may use the parametrization
\be
g = e^{x P_1} e^{uJ} e^{y P_1 + vT}
\label{ge} \ee
in which case we find,
\bs
\be
 g^{-1} \pa g = J^a T_a \;\;\;\;, \;\;\;\;
  \pa g g^{-1} = \tJ^a T_a \;\;\;\;, \;\;\;\;
g T_a g^{-1} = \o_{a}{}^b T_b
\ee
\be
J^a = (  c \pa x + \pa y , -s \pa x + y \pa u ,
 \pa u , \pa v + s y  \pa x - \frac{1}{2} y^2 \pa u )
\label{cur1}
\ee
\be
\tJ^a = (  c \pa y  + \pa x  ,  s \pa y - x \pa u ,
 \pa u , \pa v +  s x \pa y - \frac{1}{2} x^2 \pa u )
\label{cur2} \ee
\be
\o_{a}{}^b = \left( \matrix{ c & s & 0 &  s x \cr - s & c & 0  & cx +y \cr
 sy & - x - c y & 1 & -\frac{1}{2}(x^2 +y ^2 + 2c xy ) \cr
0 & 0 & 0 & 1 \cr } \right)  \;\;\;\;,\;\;\;\;
{c \equiv \cos u \atop s \equiv \sin u } \label{om}  \ee
\label{co} \es
Then, the resulting WZW action is given by
\be
I(E_2^c)  = -{k \over 4 \pi} \int \rd^2 z
( \pa x \bpa x + \pa y \bpa y + 2 c  \pa x \bpa y
+ \pa u \bpa v + \pa v  \bpa u + b \pa u \bpa u )\;\;.
\label{acte2} \ee
By rescaling the space-time coordinates, it is possible to scale out the
parameter $k$, so that we will choose $k=1$ in the remainder of the paper.
Moreover, the parameter $b$ can also be removed by coordinate transformations,
so we will set $b=0$ throughout the paper as well.

The corresponding metric in the $\s$-model description
is a conformally invariant plane gravitational wave. Changing coordinates,
\bs
\be
x = {i \over \sqrt{2}s }( e^{-iu/2} \z - e^{iu/2} \bz )
\;\;\;\;,\;\;\;\;
y = {i \over \sqrt{2}s }( -e^{iu/2} \bz + e^{-iu/2} \bz )
\ee
\be
v = -v' -\frac{bu}{2} -  \frac{1}{4s} ( \z^2 + \bz^2 - 2 c \z \bz)
\ee
\es
the geometry reads
\bs
\be
\rd S^2 = 2 \rd \z \rd \bz - \frac{1}{2} \z \bz \rd u^2 - 2 \rd u \rd v'
\ee
\be
B_{\z \bz} = iu \;\;\;\;, \;\;\;\; \Fi = {\rm const.}
\ee
\label{ba2}
\es
where we also applied a gauge transformation to the antisymmetric tensor field.
This background is of the form (\ref{pgw}), with $f(u) =0 $, $b(u)=u$,
$g(u) ={\rm const.}$, and $F(u)= \frac{1}{4}$, in agreement with (\ref{beq}).

Comparing with Table 1, we establish that this is the most symmetric plane
graviational wave, corresponding to case I, whose seven Killing vectors are in
the basis
$x^{\m} = (\z,\bz,u,v)$ of (\ref{ba2}) given by,
\bs
\be
T^\m = (0,0,0,1) \;\;\;\;,\;\;\;\;
J^{\m} = (-\z,  \bz, 0, 0 )
\;\;\;\;,\;\;\;\; M^\m = (0,0,-2,0)
\ee
\be
(P^{(i)}_{\a})^\m= ( h^{(i)}, \a h^{(i)},0,(\bz + \a \z) (h^{(i)})') \;\;\;\;,
\;\;i=1,2\;\;,\;\; \a = \pm
\ee
 \be
h^{(1)}(u) =  \cos \frac{u}{2}
\;\;\;\;, \;\;\;\; h^{(2)}(u) =  \sin \frac{u}{2}
\ee
\be
[P^{(i)}_{\a}, P^{(j)}_{\b} ] =
 \a \d_{\a \b} e^{ij}  T
\;\;\;\;, \;\;\;\; [J, P^{(i)}_{\a} ] = P^{(i)}_{-\a}
\;\;\;\;, \;\;\;\;
[M, P^{(i)}_{\a} ] = \e_{ij} P^{(j)}_{\a}
\ee
\label{e2la} \es
where we used the general expressions in (\ref{kv2}). Here, a summation over
any pair of lower and upper indices is understood, while we have used the same
notation for the Killing vectors $\xi^{\m}$ and the corresponding Lie algebra
generators $\xi^{\m} \pa_{\m}$. The form (\ref{f3}) of the metric is also
easily
obtained by taking e.g.
$h(u) =\cos(u/2)$.
As can be seen  explictly from the Lie algebra satisfied by these Killing
vectors, one
can find a subalgebra of three commuting generators.

We also remark that the corresponding exact CFT of this model was identified
as a solution of the Virasoro master equation \cite{vme},
with central charge $c_G=4$.
In particular, the stress tensor on $G=E_2^c$
\be
T_{G} = L_{G}^{ab} \xx J_a J_b \xx
\;\;\;\;, \;\;\;\;
L_{G}^{ab}  =\frac{1}{2} (\O^{-1})^{ab} + \frac{1}{2k^2} \d^a_{4} \d^b_4
 \label{e2st} \ee
is the natural generalization of the affine-Sugawara
construction \cite{bh,witkz} on $E_2^c$, satisfying the properties:
\ben
\ia all the currents
$J_a$ are primary with conformal weight $\D=(1,0)$ with respect to $T_{G}$,
\ib for any solution $L^{ab}$ of the master equation, the K-conjugate
construction
\be\tilde{L}^{ab} = L_G^{ab} - L^{ab} \;\;\;\;, \;\;\;\;
\tilde{c}= c_G - c
\label{kcon} \ee
is also a solution.
\een
A systematic approach to construct affine-Sugawara constructions on
non-semi-simple groups was given in Ref.\cite{mo}, and further exploited
\cite{sf}  to
show that the central charge of these constructions is always an integer
equal to the dimension of the Lie algebra.

\section{The Gauged WZW Model $({\bf E_2^c \ti E_2^c)/E_2^c}$}

Our aim in this paper is to compute and examine
 the geometry of the gauged WZW model
$ (E_2^c \ti E_2^c)/E_2^c$. To this end we first
 recall a result obtained in Ref.\cite{bs},
concerning  different ways of gauging a WZW model,
which will turn out to be relevant for the particular coset theory that
we wish to investigate.

Given a gauged WZW model $G/H$, for each outer automorphism
of the $H$ algebra
 there is an inequivalent way of choosing the world-sheet gauge group
in an anomaly-free way.
More precisely, let $S$ be an outer automorphism of $H$, so that
\be
S_a{}^d S_b{}^e f_{de}{}^g (S^{-1})_g{}^c = f_{ab}{}^c \;\;\;\;,\;\;\;\;
S_a{}^c S_b{}^d \O_{cd} = \O_{ab} \;\;\;\;a=1,\ldots, \mbox{dim}\,H
\label{out} \ee
where $f_{ab}{}^c$ are the structure constants of $H$ and $\O_{ab}$ the
Killing metric on $H$, or, more generally, the invariant bilinear form on
$H$ when $H$ is non-semisimple.
Then the world-sheet gauge group can be chosen to be
\be
{\cal J}_a^H = J^H_a + S_a{}^b \bar{J}_b^H
\label{wsc} \ee
where $J_a^H$ and $\bar{J}_a^H$ are the left- and right-moving world-sheet
currents of $H$ respectively. It is easy to see using the properties in
(\ref{out}) that the currents ${\cal J}_a^H$ in (\ref{wsc}) form a closed
algebra.

The corresponding action may then be written as,
\bs
\be
I(g,A) = I_{WZW}(g) + I_{gauge}(g,A)
\ee
\be
I_{gauge}(g,A) =
\frac{1}{2 \pi} \int_{\Sigma}  \rd^2 z
{\rm Tr}[A_l \bpa g g^{-1}  -  \bA_r g^{-1} \pa g + g^{-1} A_l g \bA_r  -
A_l' \bA_r ]
\ee
 \be
A_l = A_l^a T_a^{H} \;\;\;\;, \;\;\;\;
\bA_r = \bA_r^a T_a^{H} \;\;\;\;, \;\;\;\;
A_l' = A_l^a  (T')_a^H \;\;\;\;, \;\;\;\; (T')_a^H = S_a{}^b T_b^H
\ee
\label{act} \es
where the WZW action $I_{WZW}(g)$ is defined in eq.(\ref{wzw}), and
the gauge fields $A_l$ and $\bA_r$ take values in the subgroup $H$,
as indicated.
It is not difficult to check, using the Polyakov-Wiegmann identity and the
properties in (\ref{out}) that this action is invariant under the gauge
transformations
\bs
\be
g \ra h_l^{-1} g h_r
\;\;\;\;, \;\;\;\;
h_l = e^{x^a T_a^{H} }\;\;\;\;, \;\;\;\; h_r = e^{ x^a (T')_a^{H} }
\label{gt} \ee
\be
A_l \ra h_l^{-1} (A_l - \pa )h_l
\;\;\;\;, \;\;\;\;
\bA_r \ra h_r^{-1} (\bA_r - \bpa )h_r
\ee
\es
where $h_l$ and $h_r$ are elements of the subgroup $H$.

Here, the usual vector gauging corresponds to the trivial automorphism $S=1$.
The axial gauging (which is anomaly-free for abelian subgroups)
corresponds to $S = -1$, which is clearly an outer automorphism for abelian
groups. However, the result above implies that even when the subgroup is
non-abelian, the existence of non-trivial outer automorphisms
gives rise to non-equivalent ways of
gauging besides the vector gauging, which are typically of a mixed vector-axial
type. Such outer automorphisms occur for
example in $SU(n)$, $SO(2n)$ with $n \geq 3$, and $E_6$ groups (when $H$ is a
compact non-abelian subgroup). In analogy with the duality between the vector-
and axial gauging \cite{kdvv}
that was found for abelian $H$, one similarly expects a
duality between the vector and the mixed vector-axial gaugings for
non-abelian $H$.
This is indeed the case, in the
particular gauged WZW model that will be discussed below.

To obtain more general non-trivial four-dimensional string backgrounds, we
now turn to the gauged WZW model on the product group
$G=E_2^c \ti E_2^c$.
Using (\ref{ge}), the group elements of $G$ can be parametrized
as
\be
g = g_1 \ti g_2 \;\;\;\;, \;\;\;\; g_i = e^{x_i P^{(i)}_1} e^{u_i J^{(i)}}
e^{y_i P_1^{(i)} + v_i T^{(i)}}
\;\;\;,\;\; i =1,2
\label{pgg} \ee
where $T_a^{(i)},\;i=1,2$ are the generators of each of the two copies of
$E_2^c$.

The gauge group we take is $H =E_2^c$, so that a four-dimensional target
space is obtained. There is a continous embedding of $E_2^c$ in $G$, which is
given by
\bs
\be
T^H_a = T_a^{(1)} + R_a{}^b T_b^{(2)}
\ee
\be
R_a{}^b = r_a \d_a^b \;\;\;\;, \;\;\;\;
r =( \sqrt{\n}, \e \sqrt{\n} , \e , \e \n)
\;\;\;\;\e = \pm 1
\ee
\label{sgg} \es
where $\n$ is an arbitrary parameter, and $\e$ labels two distinct
sectors of the
embedding.
 The case $\e = \n =1$ corresponds to
the diagonal subgroup. Note that the same model is found,
when one leaves the levels $k_1,\;k_2$ arbitrary and one gauges
the diagonal subgroup, with the identification $\n = k_2/k_1$.

Moreover, the subgroup $H=E_2^c$ has a non-trivial outer automorphism,
\be
S_a{}^b = \s_a \d_a^b
\;\;\;\;, \;\;\;\; \s=
(1,-1,-1,-1)
\label{oute} \ee
so that, according to the general result above, we can distinguish two
different world-sheet gaugings,
the vector gauging (corresponding to $S=1$) and the
mixed vector-axial type, with $S$ given in (\ref{oute}).

To compute the action in (\ref{act}) explicitly, we
use the subgroup generators (\ref{sgg}), and the parametrization (\ref{pgg})
to obtain
\bs
\be
g^{-1} \pa g = g_1^{-1} \pa g_1 + g_2^{-1}  \pa g_2
= J_{(1)}^a T_a^{(1)} + J_{(2)}^a T_a^{(2)}
\ee
\be
\bpa g g^{-1}  =  \bpa g_1  g_1^{-1}+   \bpa g_2 g_2^{-1}
= \bar{\tJ}{}_{(1)}^a T_a^{(1)} + \bar{\tJ}{}_{(2)}^a T_a^{(2)}
\ee
\be
 g T_a^{(i)} g^{-1} = (\o^{(i)})_a{}^b T_b^{(i)} \;\;\;\;, \;\; i =1,2
\ee
\es
where the currents $J^a_{(i)}$,  $\bar{\tJ}{}^a_{(i)}$ and the matrices
$\o_{(i)}$ are given in (\ref{co}), with $x \ra x_i$, $y \ra y_i$, $u\ra u_i$
and $v \ra v_i$, $i=1,2$.
Moreover, we use the non-degenerate bilinear form (\ref{bil}) to perform
the traces over the
representation matrices, so that
\be
{\rm Tr} ( T_a^{(i)} T_b^{(j)} ) \ra \d_{ij} \O_{ab}
\;\;\;\;. \ee
Recall that we have chosen $k_1=k_2=1$ and
$b_1 = b_2=0 $ for the arbitrary
constants in the bilinear forms.

Then, using eq.(\ref{act}) we obtain
\bs
\be
I(g,A) = I_{1} (E_2^c) + I_{2} (E_2^c)
+ \frac{1}{2 \p} \int \rd^2 z [
B_a  \bar{\tJ}{}_H^a
-\bB_a
J_H^a   +  \bB_a M^{ab} B_b ]
\ee
\be
B_a \equiv A_l^b \O_{ba}  \;\;\;\;, \;\;\;\;
\bB_a \equiv \bA_r^b \O_{ba}
\ee
\be
\bar{\tJ}{}_H^a =   \bar{\tJ}{}_{(1)}^a +
\bar{\tJ}{}_{(2)}^b  \tR_b{}^a
\;\;\;\;, \;\;\;\;
J_H^a =   J_{(1)}^a +
J_{(2)}^b  \tR_b{}^a
\label{hcur} \ee
\be
M = \O^{-1}  (\o^{(1)} - S^{-1} +
R \o^{(2)} \tR - R \tR S^{-1} )
\label{matrixM}\ee
\be
 \tR_a{}^b = (\O^{-1}  R \O )^b{}_a = \hat{r}_a \d_a^b
\;\;\;\;, \;\;\;\; \hat{r} = (\sqrt{\n}, \e \sqrt{\n}, \e \n, \e)
\label{mtr} \ee
\label{acte} \es
where $I_{i}(E_2^c),\; i =1,2$ denotes the two copies of the WZW action
(\ref{acte2})
on each of the $E_2^c$ factors.

Using the form of the currents in (\ref{cur1}),(\ref{cur2}), the matrix $\o$
in (\ref{om}),
and the action (\ref{acte2}),
it is not difficult to check that the choice $\e \ra - \e $ in the
subgroup generators  (\ref{sgg})
corresponds to the coordinate transformation $u_2 \ra -u_2,\;\,v_2 \ra -v_2$
so that, without loss of generality, we can take $\e=1$ in the following.

The next step is to choose a gauge fixing, integrate out the gauge fields
and determine the background geometry by identifying the action with the
$\s$-model form in (\ref{sm}), and reading off the metric
 $G_{\m \n}$, the antisymmetric tensor $B_{\m\n}$ and the dilaton
$\Fi$.
This is done for the vector gauging in Section 5 and for the vector-axial
gauging in Section 6. The geometries that we will find are all in the
general class of plane
gravitational waves and their duals,
discussed in Section 2, and they are accompanied by a
non-constant dilaton.

Of course, we know that this model is conformally invariant to all orders,
since
there is an underlying CFT based on the $G/H$ coset construction
\cite{bh,gko}.
For completeness, we give here the form of the corresponding
stress tensor
\be
T_{G/H} = T_G - T_H \;\;\;\;, \;\;\;\; c_{G/H} = c_G - c_H = 4
\label{cos} \ee
where $T_G$ and $T_H$ are the affine-Sugawara constructions on
$G=E_2^c \ti E_2^c$
and $H=E_2^c$, respectively:
\bs
\be
T_G = \frac{1}{2} \sum_{i=1}^2
[ (\O^{-1})^{ab} +
 \d^a_4 \d^b_4 ]
\xx J^{(i)}_a J^{(i)}_b \xx
\;\;\;\;, \;\;\;\; c_G = 8
\ee
\be
T_H =
\frac{1}{2} [ (\O_{H}^{-1})^{ab} +
 \frac{1}{(1+\n)^2}\d^a_4 \d^b_4 ] \xx J^{H}_a J^{H}_b \xx
\;\;\;\;, \;\;\;\; c_H = 4
\ee
\be
J_a^H = J_a^{(1)} + R_a{}^b J_b^{(2)}
\ee
\es
with $J_a^{(i)},\;a=1,\ldots,4$ the currents of the two copies
of the affine Lie algebra
$E_2^c$.
Here, we have used the $E_2^c$ stress tensor in (\ref{e2st}) (with
$k_1 = k_2=1$ and $b_1=b_2=0$) and we have introduced the
induced bilinear form on the algebra generated by $J_a^H$:
\be
\O^H_{ab} = (1+\n) \left( \matrix{
 1 & 0 & 0 & 0 \cr
0 & 1 & 0 & 0 \cr 0 & 0 & 0
 & 1 \cr 0 & 0 & 1 & 0 \cr} \right)
\;\;\;\;.  \ee
We have verified explicitly that the coset stress tensor in (\ref{cos})
sastisfies
the Virasoro master equation, as it should according to the K-conjugation
property given in (\ref{kcon}).

Note that the central charge of the construction is exactly
 equal to 4, so that, in particular, the central charge deficit appearing in
the
one-loop dilaton beta function (\ref{betadil}) vanishes.

We finally remark that the gauged WZW model should have a remaining chiral
$U(1)$ current, since
\be
{\cal J} = J_4^{(1)} - J_4^{(2)}
\ee
is a dimension $\D=(1,0)$ operator of the coset construction in (\ref{cos}).
In fact, this chiral symmetry, is the origin of the existence of a null
Killing vector in the geometries we will find below.

\section{The Vector-Gauged Model}

In this section, we discuss the evaluation of the action (\ref{acte}) for
the
vector gauging, corresponding to $S=1$.
 First, we find a gauge-fixing by
studying the transformations in (\ref{gt}) which show that we may choose
\be
x_1 = {x_2 \over \sqrt{\n} } \equiv r
\;\;\;\;, \;\;\;\;
y_1 =  y_2 =0
\;\;\;\;. \ee
Note that it is only possible to eliminate three from the eight degrees of
freedom, so that naively a five dimensional target space in the gauged
model is expected. However, when integrating out the
gauge fields in (\ref{acte}) we also find an additional
constraint, which eliminates a fourth degree of freedom.

In fact, the matrix $M$ is singular in this case with
\be
M_{a}{}^ 3 = M_3{}^a = 0 \;\;\;\;, \;\;a= 1,2,3,4
\ee
so that when integrating out the fields $B_3$ and $\bB_3$ one finds the
constraints
\be
 \bar{\tJ}{}_{H}^3    =
 J_H^3   =0 \;\;\;\;.
\label{cons} \ee
Then, substituting the explicit form of the currents in (\ref{hcur}),
(\ref{cur1}), (\ref{cur2})
 and using $\tR$ from  (\ref{mtr}), we have
\be
\bpa (u_1 + \n u_2 ) = \pa (u_1 + \n u_2 ) =0
\;\;\;\;. \ee
The general solution is
\be
u_1 = u(1- \k ) + \r
\;\;\;\;, \;\;\;\;
u_2 = -u(1+ \k) + \r
\ee
where we have defined
\be
\k \equiv {1- \n \over 1 + \n }
\label{kap} \ee
while $\r$ is an arbitrary integration constant.
Moreover, for the remaining two variables it is useful to
perform the change of basis:
\be
v_1 = \frac{1}{2} \left( \th -{ 2 \over (1-\k)(1+\k)^2} v \right)
\;\;\;\;, \;\;\;\;
v_2 = \frac{1}{2} \left( {1- \k \over 1+\k}  \th + { 2 \over (1-\k)(1+\k)^2}
v \right)
\;\;\;\;.
\ee

Then we can integrate the remaining gauge fields, and obtain
\be
I_{\mbox{vector}} = I_{1}(E_2^c) + I_{2}(E_2^c)
+  {1 \over 2 \p} \int \rd^2 z
 \sum_{a,b= 1,2,4} \bar{\tJ}{}_{H}^a
   (\hat{M}{}^{-1})_{ab}
J_H^b
\ee
where the non-singular reduced $3\ti 3$ matrix $\hat{M}$ is
\bs
\be
\hat{M} = \left(\matrix{
c_+ - 1 + \n (c_- -1) & s_+ -\n s_- & ( s_+ -\n s_-)r \cr
-s_+ + \n s_- & c_+ -1 + \n (c_- -1) & (c_+ + \n c_- )r \cr
0 & -(1 + \n)r & -(1 + \n)\frac{r^2}{2}  \cr } \right)
\ee
\be
{\rm det} \hat{M} = -4 r^2 s^2 \frac{(1-\k)}{ (1+ \k)^2}
\;\;\;\;, \;\;\;\; c_{\pm} \equiv \cos( u( 1 \mp \k) \pm \r )
\;\;\;\;.
\ee
\es
Moreover, as a result of the integration, we
also get a non-trivial dilaton field
$\Fi= -\frac{1}{2}  \ln({\rm det}\hat{M})$.

The final result for the
background geometry (up to a common multiplicative factor) is
\bs
\be
dS^2 =\frac{1}{s^2} \left[  F_+(u,\r) \rd r^2 +
\frac{4}{r} (\k s \tc_\r - c \ts_\r  )\rd r  \rd \th
+\frac{F_-(u,\r)}{r^2} \rd \th^2 \right] - 2 \rd u \rd v
\ee
\be
B_{\m \n } = 0 \;\;\;\;,\;\;\;\; \Fi = -\ln(r s )
\ee
\label{geov} \es
where we defined
\bs
\be
c = \cos u \;\;\;\;,\;\;\;\; s =\sin u\;\;\;\;,\;\;\;\;
\tc_\r  = \cos( \k u - \r) \;\;\;\;,\;\;\;\;\ts_\r  = \sin( \k u - \r)
\label{csdef} \ee
\be
F_{\pm} (u,\r)= 1 + \k^2 + (1-\k^2) c^2 \pm 2 (c \tc_\r + \k s \ts_\r)
\;\;\;\;.
\label{fdef}\ee
\label{def} \es
The geometry (\ref{geov}) depends on the two arbitrary parameters $\k$
and $\r$.

Comparing with the general background (\ref{geod}) dual to the
plane wave, we see that (\ref{geov}) is precisely of this form (up to
a constant rescaling $r \ra r/(1-\k^2) $), with the identifications
\be
h(u) = {s \over F_- (u,\r)^{1/2} }
\;\;\;\;,\;\;\;\;
 l(u) = {2( c \ts_\r - \k s \tc_\r ) \over (1-\k^2) s^2 }
\;\;\;\;, \;\;\;\; g(u) =  F_- (u,\r)^{1/2}
\label{rel1} \ee
where we used the relation
\be
F_-(u,\r) F_+(u,\r) - 4 ( c \ts_\r - \k s \tc_\r )^2 = (1-\k^2)^2 s^4
\;\;\;\;. \ee
It follows from the discussion of Section 2, that to check the one-loop beta
functions, we may use the results in (\ref{rel1}) to compute
$F(u) = h(u)''/h(u)$ and
$b(u)'=[ h(u)^2 l(u)]'/h(u)^2$
(see (\ref{skv}) and (\ref{bkrel})). Using also $g(u)$, one can
verify that the condition (\ref{beq}) is
indeed satisfied as it should.

The Ricci tensor for the metric can be obtained using (\ref{geoq2})
and (\ref{rel1})
and, in particular,
 the curvature scalar of the metric is
\be
R =- { 4 F_-(u,\r) \over (1-\k^2)^2 r^2 s^2 }
\ee
which has singularities at $r=0$ and $u= n \pi, \; n \in \Z$.

\section{The Mixed Vector-Axial Gauged Model}

In this section, we evaluate the action (\ref{acte}) for the mixed
vector-axial gauging,
which corresponds to the non-trivial outer automorphism $S$ in (\ref{oute}).
First, by studying the gauge transformations in
(\ref{gt}),
we make the gauge-fixing choice:
\be
x_1 = {x_2 \over \sqrt{\n} } \equiv x
\;\;\;\;,\;\;\;\;
y_1 = - {y_2 \over \sqrt{\n} } \equiv y
\;\;\;\;,\;\;\;\;
u_1 = -u_2 \equiv u
\;\;\;\;, \;\;\;\;
v_1 = - {v_2 \over \n} \equiv v
\;\;. \ee
Note that, in contrast to the vector gauging of the previous section, the
gauge fixing now eliminates four degrees of freedom, as expected generically.
Indeed, in this case the matrix $M$ in eq.(\ref{matrixM}) is non-degenerate, so
that no additional constraints arise from integrating out the gauge fields.

The resulting action is
\be
I_{\mbox{vector-axial}} = I_{1}(E_2^c) + I_{2}(E_2^c) +
\frac{1}{2 \p} \int \rd^2 z \bar{\tJ}_H^a (M^{-1})_{ab} J_H^b
\ee
where the matrix $M$, in terms of
the parameter $\k$ defined in (\ref{kap}),
takes the form:
\bs
\be
M =\frac{2}{1+\k} \left( \matrix{  c-1 & \k s & 0 & \k  s x   \cr
-\k s & c +1  &  0 & cx + \k y  \cr
0 & 0 & 0  & 2 \cr
 s y & -x   + \k c y & 2 & -\frac{1}{2} (x^2 + y^2)
- c x y   \cr }
\right)
\ee
\be
{\rm det} M = {64 (1-\k) s^2 \over (1+\k)^3}
\;\;\;\;. \ee
\es
After some algebra, one obtains the
following background geometry
(up to a multiplicative factor):
\bs
$$
\rd S^2 = {2\over 1-c}  [ 1+ \k^2 + (1-\k^2) c] \rd x^2
+ {8 \k \over 1-c} \rd x \rd y + {2\over 1-c} [ 1+\k^2 - (1-\k^2) c] \rd y^2
$$
$$
+ {\k \over 2(1+c)} [ (1 + \k^2 +(\k^2-1) c) \k x^2 + 4 x y +
 \k (5 - 3 \k^2 +c (1-\k^2)) y^2 ]  \rd u^2
$$
$$
+ 4 (1-\k^2)^2 \rd u \rd v
+
{2 \k \over s} [-2 \k x  + (\k^2 -3 + (\k^2-1) c )y ] \rd x \rd u
$$
\be
-
{2 \over s} [ (1 + \k^2 +(\k^2-1) c) \k x + 2 (1 -c (1-\k^2)^2) y ] \rd y \rd u
\label{m1} \ee
\be
\Fi = - \ln s
\ee
\be
B_{\m \n} = \pa_{\m} \L_{\n} - \pa_{\n}  \L_{\m}
\;\;\;\;, \;\;\;\; \L_x =\L_y = 0  \;\;\;\;,\;\;\;\;
\L_u = \frac{1}{2} (1-\k^2) y(y + \k x) {s \over 1+c}
\ee
\es
so that the antisymmetric tensor field is pure gauge and can be
discarded.

After a tedious but straightforward calculation, whose details can be
found in Appendix A, one finds that the
metric (\ref{m1}) can be transformed  into the
two alternate forms
\bs
\be
\rd S^2 = 2 \rd \z \rd \bz - 2( f(u) \z^2 + \fb(u) \bz^2 + F(u) \z \bz) \rd u^2
- 2 \rd u \rd v
 \ee
\be
f (u)= \frac{1}{2s^2} ( c - i \k s ) e^{i \k u} \;\;\;\;,\;\;\;\;
F (u)= -\frac{1}{s^2}
\label{abv} \ee
\label{geova} \es
or
\be
\rd S^2 = \frac{1}{s^2} [ F_-(u,0) \rd x^2 + 4(c \ts_0 - \k s \tc_0) \rd x
\rd y
+ F_+(u,0) \rd y^2 ] - 2 \rd u \rd v
\label{vaba2} \ee
where in the last form we have used the definitions $\tc_\r,\;\ts_\r$ and
$F_{\pm}(u,\r)$ in (\ref{def}) at $\r =0$. Note that this class of backgrounds
depends on one continuous parameter $\k$.

Comparison of (\ref{geova}) with Table 1, shows that this is a generic plane
gravitational wave with five Killing vectors,
 corresponding to case IV. It is easy to check that
the relation (\ref{beq}) is obeyed using $F(u)=(\sin u)^{-2} $, $g(u) = \sin
u$,
and $b(u)=0$, so that the
background is conformally invariant to all orders.
The explicit form of the non-vanishing component of the Ricci tensor and the
curvature scalar are
\be
R_{uu }= - \frac{2}{s^2}
\;\;\;\;, \;\;\;\; R =0
\ee
where we used eq.(\ref{geoq}) and $F(u)$ in (\ref{abv}).

As an example, we have computed the five Killing vectors when
$\n =1\;(\k=0) $. In the
basis $x^{\m}=(\z,\bz,u,v)$ of (\ref{geova}) they are
given by
\bs
\be
T^{\m} = (0,0,0,2)
\ee
\be
(P^{(i)}_{\a})^{\m}  =
(f_{\a}^{(i)},\a f_{\a}^{(i)},0,( \bz +\a \z) (f_{\a}^{(i)} )' )
\;\;\;\;,\;\; i=1,2\;\;\;,\;\;
 \a = \pm
\ee
\be
f_{\a}^{(1)}(u) = {\a s \over 1 - \a c} \;\;\;\;, \;\;\;\;
f_{\a}^{(2)}(u) = 2  - \a { u s \over  1- \a c }
\ee
\es
where we used the general formula in (\ref{kv}). This set of Killing vectors
satisfies the commutation relations $ [P_{\a}^{(i)},P_{\b}^{(j)}] =
 \a  \d_{\a \b}  \e^{ij} T$, which defines a subalgebra of the
7-dimensional algebra in (\ref{e2la}).

\section{The Dual Backgrounds}

As announced in Section 4, one expects the different geometries obtained by
the vector and mixed vector-axial gauging of the $(E_2^c \ti E_2^c)/E_2^c$
model, to be related by duality. In this section, we establish this connection.

Starting with the geometry (\ref{geov}), the dual background obtained using
the isometry in the $\th$-direction, can be read off immediately from
(\ref{f3}) and the identifications in (\ref{rel1}), since we already
showed that (\ref{geov}) is dual to a plane gravitional wave.
In fact, an $O(1,1)$ duality in the $\th$-variable
results in
\bs
\be
dS^2 =\frac{s^2}{F_- (u,\r)} [ \rd r^2 +
r^2 \rd \th^2 ] - 2 \rd u \rd v
\ee
\be
B_{r\th}= {2r(c \ts_\r  - \k s \tc_\r ) \over (1-\k^2) F_-(u,\r) }
\;\;\;\;, \;\;\;\;
\Fi = -\frac{1}{2} \ln(F_-(u,\r)  )
\;\;\;\;. \ee
\label{vecd} \es
For completeness, we also give the alternate form
\bs
\be
dS^2 = 2 \rd \z \rd \bz -2 F(u) \z \bz \rd u^2 - \rd u \rd v
\label{vecd2} \ee
\be F(u) = \frac{h(u)''}{h(u)} \;\;\;\;, \;\;\;\;
h(u) = {s \over F_-(u,\r)^{1/2} }
\ee
\be
B_{\z \bz} = \int \rd u
\left({2i (c \ts_{\r} - \k s \tc_{\r} )  \over (1-\k^2) F_-(u,\r) } \right)'
{ F_-(u,\r) \over s^2 } \;\;\;\;, \;\;\;\; \Fi = -\frac{1}{2} \ln(F_- (u,\r)  )
\ee
\es
showing that this is a gravitational wave of the type II in Table 1, with
six Killing vectors. One can check conformal invariance by verifying
the relation (\ref{beq}), and
relevant geometric quantities can be easily read off from
(\ref{geoq}).

As an example, we have computed again the six Killing vectors
when $\n=1$ ($\k=0$). In the basis $x^{\m} = (\z,\bz,u,v)$ of
 (\ref{vecd2}), these are given by
 \bs
\be
T^{\m} = (0,0,0,2)
\;\;\;\;, \;\;\;\; J^{\m} =(-\z,\bz, 0, 0 )
\ee
\be
(P^{(i)}_{\a})^{\m}  = (h^{(i)},\a h^{(i)},0,(\bz + \a \z) (h^{(i)})' )
\;\;\;\;, \;\; i=1,2 \;\;, \;\; \a = \pm \ee
\be
h^{(1)}(u)= {s \over 1-c} \;\;\;\;, \;\;\;\; h^{(2)}(u) = 2 - { us \over 1-c}
 \ee
\es
where we used (\ref{skv}). The commutation relations of the corresponding
generators $P_{\a}^{(i)},\;T,\;J$ define a six-dimensional subalgebra of
the Lie algebra in (\ref{e2la}): $[J, P^{(i)}_{\a} ] = P^{(i)}_{-\a}$,
$[P^{(i)}_{\a}, P^{(j)}_{\b} ] = \a \d_{\a \b}  \e^{ij} T$.

Turning to the dual of the vector-axial gauged model, we note that we now have
three commuting isometries at our disposal, generating an $O(3,3)$ duality.
To establish the connection with the vector-gauged model,
it suffices, however, to restrict to the duality transformations
corresponding to the space-like Killing symmetries.

Looking at (\ref{vaba2}), we see that we can either compute the dual
with respect
to the $O(2,2)$ transformations in the $(x,y)$-coordinates or we can
take the dual with respect to
an isometry formed by an arbitrary linear combination of $x$ and $y$.
In the first case, one finds that the metric is self-dual. In the second case,
we first apply the rotation
\be
x' = \cos(\r/2) x + \sin(\r/2) y  \;\;\;\;, \;\;\;\;
y' = -\sin(\r/2) x + \cos(\r/2) y
\label{ct} \ee
which leaves the metric (\ref{vaba2}) of the same form,
with the substitutions
\be
\tc_0  \ra \tc_\r  \;\;\;\;, \;\;\;\;
\ts_0  \ra \ts_\r  \;\;\;\;, \;\;\;\;
F_-(u,0) \ra F_-(u,\r) \;\;\;\;. \ee
Of course, the parameter $\r$ introduced in this way is unphysical at this
point.
 However, the dual metric
 with respect to the $x'$-isometry coincides with
the background
(\ref{vecd}) obtained as the dual of the vector-gauged model, which
has a non-trivial $\r$-dependence.

We recall here that the $\r$-dependence of the dual background (\ref{vecd})
originates from the constraint (\ref{cons}) when viewed as the dual of the
vector-gauged model. On the other hand, when viewed as the dual of the
vector-axial gauged model, its origin lies in the coordinate transformation
(\ref{ct}).

Hence, we have established that both the vector and the vector-axial gauged
models are mapped onto the same geometry with duality tranformations.
Note that these transformations
modify the number of isometries of the background at each stage of their
application.

\section{Conclusions}

We have discussed the most general class of conformally invariant $\s$-models
corresponding to strings propagating in a background of plane graviational
waves in four dimensions. These backgrounds can be classified according to
their number of Killing symmetries, ranging from the generic number five to
the most symmetric case with seven Killing vectors.
In all cases the Cartan subgroup consists of three commuting isometries,
two space-like and one null.

For a subset of the isometries of these plane graviational wave backgrounds
the dual geometries were computed, showing in particular, that either
other plane wave solutions are obtained or geometries with non-zero curvature
scalar. The latter correspond to new singular solutions of Einstein's
equations with
non-trivial matter, possessing one (``rotational'')
space-like and one null Killing vector.

The WZW model on the non-semi-simple group $E_2^c$ provides an explicit
conformal field theoretic realization of the most symmetric plane wave.
To find realizations of more general conformally invariant
plane graviational
waves and their duals, we have considered the gauged WZW model on
$(E_2^c \ti E_2^c)/E_2^c$. This coset contains an arbitrary parameter
corresponding to the continuous embedding of the subgroup $H=E_2^c$ in the
product group $E_2^c \ti E_2^c$. Moreover, two inequivalent world-sheet
gaugings
for this subgroup were found, the usual vector-gauging, and a mixed
vector-axial
gauging which arises due to a non-trivial outer automorphism of $H$.

The vector gauging leads to a singular
geometry (with non-zero scalar curvature)
which is dual to a plane graviational
wave background,
while the mixed vector-axial gauging gives rise to a plane graviational
wave with five isometries. These distinct backgrounds were
shown to be related by
duality transformations, as one expects by analogy with the axial-vector
duality
for abelian subgroups. The results are summarized in the
following diagram
$$
{E_2^c \ti E_2^c \over E_2^c}\;: \;\;\;\;
\matrix{ \bullet\; \mbox{vector gauging}
\;\;\;\;\;\;\;\;\;\;\;\;\;\;\;\;\;\;\;\;
\cr \;\; \Rightarrow \mbox{singular geometry} \;\;\;\;\;\; \;\;\; \cr
\mbox{(2 isometries)}
\cr \bullet \;
\mbox{mixed vector-axial gauging}
\cr \;\; \Rightarrow
\mbox{plane graviational wave } \cr
\mbox{(5 isometries)}\cr } \;\;\;\;
{\searrow \atop \nearrow} \matrix{ \;\;\;\; \cr \cr  \;\; \mbox{dual
geometry}\Rightarrow
\;\;\;\;\;\;\;\;\;\;\;\; \cr
        \mbox{plane gravitational wave} \cr
        \mbox{(6 isometries)} \cr }
$$
where the dual geometry was obtained via $O(1,1)$ duality transformations. For
the vector gauged model, we employed the rotational isometry, while for the
mixed vector-axial gauged model, a one-parameter linear combination of two
translational isometries was used. All three  geometries displayed in
the above diagram exhibit a different number of Killing vectors, as indicated.

It is an open problem to investigate whether this duality symmetry between
the vector and vector-axial gauged model
is an exact symmetry of the corresponding conformal field theories.
It is also interesting to derive the string excitations around these
backgrounds and study their interactions by computing physical quantities, such
as the partition function and correlation functions.

\section*{Acknowledgements}
We are grateful to P. Mazur for motivating this investigation and for his
contribution at the early stages of this work.
We also thank C. Bachas, A. Kehagias, E. Kiritsis and N. Mohammedi
for helpful discussions.

\newpage
{\bf \noindent Appendix A: \\
Coordinate transformations in the vector-axial
gauged model}
\bigskip

\noindent
In this Appendix, we give the coordinate transformations involved in
transforming the metric (\ref{m1}) obtained directly from the vector-axial
gauged WZW model to the simpler forms in (\ref{geova}) and (\ref{vaba2}).
These computations were performed using the algebraic
manipulation program Maple, and we will give the main logical steps,
while avoiding lengthy intermediate results.

First, we make the following basis transformation in (\ref{m1}),
$$
x = \frac{1}{1-\k^2} (x_1 - \k y_1) \;\;\;\;, \;\;\;\;
y = \frac{1}{1-\k^2} (y_1 - \k x_1)
\eqno(A.1a)
$$
$$
v = v_1 + \frac{1}{s} [ \k^2 c x_1^2 + \k ( 1-c) x_1 y_1 + (c-1) y_1^2 ]
\;\;\;\;
\eqno(A.1b) $$
and the metric takes the simpler form,
$$
\eqalign{
\;\;\;\;
\rd S^2 = &  2 { 1+c \over 1-c} \rd x_1^2 + 2 \rd y_1^2 -4 \k {s \over 1-c} y_1
\rd x_1 \rd u
\cr
& + \frac{1}{2s^2} [\k^2 ( 1+c)^2 x_1^2
 + (1-c) ( 4 + \k^2 (1+c) ) y_1^2 ] \rd u^2
-2 \rd u \rd v_1
\;\;\;\;. \cr}
\eqno(A.2) $$
Next, we let
$$
x_1 = x_2 + a(u) y_2 \;\;\;\;, \;\;\;\; y_1 = y_2 + b(u) x_2
\;\;\;\;, \;\;\;\; 1 - a(u) b(u) \neq 0
\eqno(A.3a) $$
$$
v_1 = v_2 + d(u) x_2 y_2
\eqno(A.3b) $$
and we determine $a(u),\;b(u),\;d(u)$
such that the new metric does not contain terms of the form
$\rd x_2 \rd y_2 $,
 $y_2 \rd x_2 \rd u$ and
 $x_2 \rd y_2 \rd u$. These three conditions result in the
equations
$$ (1-c) a(u)' + a(u) + \frac{\k}{2} [ (1+c) a(u)^2 + (1-c) ] = 0
\eqno(A.4a) $$
$$ b(u) = -{1+c\over 1-c} a(u)
\eqno(A.4b) $$
$$ d(u) = {2 \over 1-c} [(1+c) a(u)' -  \k s]
\;\;\;\;.
\eqno(A.4c) $$
The differential equation (A.4a) is of the Riccati type, and using
the particular solution $a(u) = i [(1-c)/(1+c) ]^{1/2}$ (wich corresponds to
a singular transformation in (A.3a)) it is not difficult
to find the one-parameter solution
$$ a(u) = -i \frac{1-c}{s} \left( { \t e^{i \k u} - 1
 \over \t
e^{i \k u}
+ 1 } \right)
\eqno(A.5) $$
where $\t $ is an arbitrary constant. Choosing $\t= 1$ one finds:
$$ a(u) = {1-c \over s} \tan (\k u /2) \;\;\;\;, \;\;\;\;
b(u) = -{1+c \over s} \tan (\k u /2)
\eqno(A.6a) $$
$$ d(u) = {1 \over 1-c} \left(  {\k s \over \cos^2(\k u /2)} +
2 \tan(\k u/2) - 2\k s
\right)
\;\;\;\;.
\eqno(A.6b) $$

Thus, using (A.3) we obtain a metric of the form:
$$ \rd S^2 = g_1(u)  \rd x_2^2 + g_2(u) \rd y_2^2 +
+ 2[f_1 (u) x_2 \rd x_2  + f_2(u) y_2 \rd y_2 ] \rd u
\eqno(A.7a) $$
$$
+ [p_1 (u) x_2^2 + p_2(u) y_2^2 + p_3 (u)x_2y_2 ] \rd u^2 - 2 \rd u \rd v_2
$$
$$ g_1 (u)= 2{ 1+c \over (1-c)\cos^2(\k u/2)}
\;\;\;\;, \;\;\;\;
g_2 (u)= 2{ 1 \over \cos^2(\k u/2)}
\eqno(A.7b) $$
$$ f_1(u)={s \over \cos^2(\k u/2) (1-c)^2 }
[  \k s (1 + 2 \cos^2(\k u/2) ) \tan(\k u/2)
 - 2  \sin^2(\k u/2)]
\eqno(A.7c) $$
$$
f_2(u)={1 \over \cos^2(\k u/2) s }
[  \k s (1 - 2 \cos^2(\k u /2) ) \tan(\k u/2)
 + 2 \sin^2(\k u/2)]
\eqno(A.7d) $$
where we omit listing the functions
 $p_i(u),\;i=1,2,3$ since they are not needed for the final coordinate
transformation.
Then, we set
$$ x_2 = \frac{i}{ \sqrt{2 g_1(u)}} ( \z - \bz)
\;\;\;\;, \;\;\;\;
y_2 = \frac{1}{ \sqrt{2 g_2(u)}} ( \z + \bz)
\eqno(A.8a) $$
$$ v_2 = v - \frac{1}{8}\left[ {(2f_1(u) - g_1(u)')\over g_1(u)} (\z - \bz)^2
 -{ (2 f_2(u) - g_2(u)')\over g_2(u)} (\z + \bz)^2 \right]
\eqno(A.8b) $$
which leads to the metric given in (\ref{geova}).

To show the equivalence with the alternate form in (\ref{vaba2}), we start with
that metric and discuss how to get back to the form (\ref{geova}).
First, we make the following transformation in (\ref{vaba2}),
$$ x =\frac{i}{1-\k^2} ( \a_2(u) \z' - \ba_2(u) \bz' )  \;\;\;\;, \;\;\;\;
y =\frac{i}{1-\k^2} ( -\ba_1 (u) \z' + \a_1 (u)\bz' )
\eqno(A.9a) $$
$$ \a_1(u) = { 1 \over s [2 F_+(u,0)]^{1/2} }( (1-\k^2) s^2 + 2i
(c \ts_0 - \k  s \tc_0 ))
 \;\;\;\;,\;\;\;\;
\a_2 (u)= {i [2 F_+(u,0)]^{1/2} \over 2s }
\eqno(A.9b) $$
$$ v = v_1 + \frac{1}{4}( q(u) \z'{}^2 + \bar{q} (u) \bz'{}^2 )
\eqno(A.9c) $$
$$ q(u) = -\frac{2}{s F_+(u,0)}
(2c + \tc_0 + c^2 \tc_0 + 2\k c s \ts_0 - \k^2 s^2 \tc
+i[(1+\k^2) s^2 \ts_0 + 2 \k s(1+ c \tc_0) ]  )
\eqno(A.9d) $$
which was found by determining $\a_1,\;\a_2$ such that
$g_{mn} = 2 \ba_{(m} \a_{n)} $ in the subspace spanned by $x,y$
(this corresponds to taking the inverse of the
transformation (\ref{bt})).
The metric (\ref{vaba2}) then takes the form
$$ \eqalign{ \;\;\;\;\;\;\;\;\;
 \rd S^2 = & 2 \rd \z' \rd \bz' + i {\rm Im}[q(u)]
(\bz' \rd \z'  - \z' \rd \bz') \rd u \cr
& -2[ \hat{f}(u) \z'{}^2 + \bar{\hat{f}}(u) \bz' {}^2 + \hat{F}(u) \z' \bz' ]
\rd u^2
- 2\rd u \rd v_1 \cr}
\eqno(A.10a)$$
$$ \hat{f}(u) = |f(u)| e^{ i \th_f (u)}
\;\;\;\;, \;\;\;
|f(u)|^2 = (c^2 + \k^2 s^2)/(4 s^4)
\eqno(A.10b) $$
where the
angle $\th_f(u)$ and function $\hat{F}(u)$ are complicated functions of $u$,
which we do not list here.

Finally, we perform the rotation
$$ \z' = e^{i \phi(u)/2 } \z \;\;\;\;, \;\;\;\;
\bz' = e^{-i \phi(u)/2 } \bz
\eqno(A.11) $$
so that the metric becomes
$$ \rd S^2 = 2 \rd \z \rd \bz + i( {\rm Im}[q(u)] - \phi(u) )
(\bz \rd \z  - \z \rd \bz)
 \rd u
\eqno(A.12) $$
$$
-2[ |f(u)| e^{i\th_f(u) + i\phi(u)  } \z^2 +
|f(u)| e^{-i\th_f(u) - i \phi(u)}  \bz^2 + (\hat{F}(u) -
\frac{1}{4} (\phi(u)')^2 ) \z \bz ] \rd u^2
- 2\rd u \rd v_1
$$
We determine the phase $\phi(u)$ by requiring the off-diagonal terms
$(\bz \rd \z  - \z \rd \bz) \rd u $ to vanish, so that
$\phi(u)' =  {\rm Im}[q(u)]$ .
This differential equation can be simplified by first shifting out the phase
$\th_f(u)$, which leads to
$$  \phi = \tilde{\phi} - \th_f
\;\;\;\;, \;\;\;\;
\tilde{\phi}(u)' = {\k(1-\k^2)   s^2 \over c^2 + \k^2 s^2}
\eqno(A.13) $$
with solution:
$$ \tilde{\phi}(u)
= \arctan \left( { -c \ts_0 + \k s \tc_0 \over c \tc_0 + \k s \ts_0 } \right)
\;\;\;\;.
\eqno(A.14) $$
Inserting this expression in the metric (A.12), and subsequently letting
$u\ra -u$, we find exactly the form (\ref{geova}).

\end{document}